\documentclass[12pt]{article}
\usepackage{physics}
\usepackage{amsmath}
\usepackage{amssymb}
 \usepackage{notoccite}
 \usepackage{esint}
 \usepackage{graphicx}
 \usepackage{caption}
\usepackage{subcaption}

\begin{document}
\input epsf
\def\be{\begin{equation}}
\def\bea{\begin{eqnarray}}
\def\ee{\end{equation}}
\def\eea{\end{eqnarray}}
\def\d{\partial}
\def\la{\lambda}
\def\eps{\epsilon}
\def\l#1{ \label{#1}}
\def\half{ { 1\over 2}}

\def\nono{ { \nonumber}}


\begin{center}
\Large{\bf $1/2$ BPS Structure Constants and Random Matrices}

\vspace{ 20mm}

\normalsize{Adolfo Holguin }

\vspace{10mm}

\normalsize{\em  Department of Physics, University of California Santa Barbara, CA 93106}

\vspace{0.2cm}

\end{center}

\vspace{10mm}

\begin{abstract}
\medskip
We study three point functions of half BPS operators in $\mathcal{N}=4$ super Yang-Mills theory focusing on correltors of two of the operators with dimension of order $\Delta\sim N^2$ and a light single trace operator. These describe vacuum expectation values of type IIB supergravity modes in LLM backgrounds that do not necessarily preserve the same symmetries as the background solution. We propose a class of complex matrix models that fully capture the combinatorics of the problem, and describe their solution in the large $N$ limit.  In simple regimes when the dual description is in terms of widely separated condensates of giant gravitons we find that the models are solvable in the large $N$ and can be approximated by unitary Jacobi ensembles; we describe how these distributions are reproduced in the dual bubbling geometry picture for large droplets. In the case of two eigenvalue droplets the model is exactly solvable at finite $N$. As a result we compute all half-BPS structure constants of heavy-heavy-light type, and reproduce the formulas found via holographic renormalization in the large $N$ limit. We also comment on structure constants of three heavy operators.  
\end{abstract}

\newpage

\section{Introduction}
\renewcommand{\theequation}{1.\arabic{equation}}
\setcounter{equation}{0}

The study of baryonic operators in large $N$ gauge theories is an old subject \cite{Witten:1979kh} that has received renewed attention in the context of holographic field theories \cite{Jiang:2019xdz, Chen:2019gsb, Berenstein:2022srd}. Such operators are extremely interesting from the point of view of the large $N$ expansion, since they correspond to heavy non-perturbative objects that are not very well captured by the conventional t' Hooft expansion. On physical grounds one expects that such heavy objects modify the physics at the semi-classical level, and that one should attempt to approach the problem from a point of view where one treats the dynamics of the many constituents of the object in terms of a simpler collective coordinates \cite{Gervais:1974dc, Gervais:1975yg}. This is well understood in string theories; heavy objects can lead to non-trivial boundary conditions for strings, or in some cases deform the target space geometry which the string probes.  For this reason such an approach is essential for understanding how gravitational physics arises from large $N$ models. In examples of the AdS/CFT correspondence \cite{maldacena1999large} these ideas have sharp realizations in terms of \textit{giant gravitons}, and in suitable limits, non-trivial supergravity backgrounds like \textit{bubbling geometries} and \textit{black holes}. Given that maximally supersymmetric Yang-Mills theory is expected to be fully-fledged theory of quantum gravity, a particularly interesting question to address is how $\mathcal{N}=4$ SYM resolves the many puzzles of gravitational theories, in particular the physics of black holes. At the moment such questions are out of reach and they lie in regions of parameter space where current non-perturbative techniques such as integrability are expected to fail. One particular fruitful approach has been to concentrate on observables which are protected by supersymmetry in order to test and develop tools, and the half-BPS sector of asymptotically AdS type IIB supergravity and $\mathcal{N}=4$ SYM is perhaps the simplest non-trivial toy model.

The spectral problem in this sector of the $U(N)$ theory was solved by the work of Corley, Jevicki, and Ramgoolam \cite{Corley:2001zk}; half-BPS operators are Schur functions and their structure constants are given by multiplicities of representations of the unitary group
\begin{equation}
    \left< \mathcal{O}_{R_{3}}(\bar{Z}) \mathcal{O}_{R_{2}}(\bar{Z}) \mathcal{O}_{R_{1}}(Z) \right>= C_{R_1 R_2 R_3} f_{R_1},
\end{equation}
where $C_{R_1 R_2 R_3}$ are Richardson-Littlewood coefficients and $f_R$ is the norm of the operator $\mathcal{O}_R$. Although this solves the problem in principle and combinatorial algorithms exist which generate these coefficients for a fixed value of $N$ it is unclear how the asymptotics of these coefficients are reflected in the corresponding supergravity solutions. Since these numbers appear naturally in the study of the intersection theory of Grassmannians, a natural expectation is that there is an alternative description for such calculations involving only geometric data coming from the gauge group of the theory. Another issue is that most results in the existing literature on structure constants of half-BPS operators either focus on single trace operators, or in operators preserving the same supersymmetries, or rely heavily on the free fermion description of model. Holographic computations of one point-functions in half-BPS backgrounds have also been studied in generality, but explicit calculations are limited to maximally charged operators which are charged under the same symmetry as the background and to operators of low dimensions. So an important step towards understanding very heavy operators in $\mathcal{N}=4$ SYM is to develop tools that can tackle problems of this kind for generic BPS operators in the large $N$ limit. Such tools have been developed recently for operators of dimension $\Delta \sim N$ \cite{Jiang:2019xdz, Chen:2019gsb, Yang:2021kot, Lin:2022gbu, Holguin:2022zii, Carlson_2023} and in this paper we extend this to operators of dimensions that scale as $N^2$. We show that the computation of very generic three-point functions of half-BPS operators can be packaged in a large family complex matrix model of matrices valued on a Grassmannian. Although we mostly focus on the $U(N)$ theory, our results generalize readily to orthogonal and symplectic gauge groups. For simple observables, such as set-ups involving a single stack of AdS giant gravitons, the corresponding matrix ensemble is a unitary Jacobi ensemble, while for more generic observables the matrix model cannot be easily reduced to integrals over eigenvalues. At large $N$, we find that the saddle point equations simplify the calculation significantly allowing us to either reduce the integrals to sums over integrals over eigenvalues, where each term in the sum is labelled by a permutation. The average density of eigenvalues is universal and is given by the well-known Marchenko-Pastur distribution, which appears as the Poisson distribution of noncommutative probability theory \cite{Livan_2018}
\begin{equation}
  \rho(z)=\frac{\sqrt{(z_+-z)(z-z_-)}}{2\pi z}\;.
\end{equation}
Gaussian matrix integrals have been used extensively in the study of the combinatorics of half-BPS correlators in the past \cite{ berenstein2004toy, Takayama:2005yq} and in other contexts \cite{Drukker:2000rr, Pestun:2007rz}; the matrix models we study on the other hand describe large deviations from the vacuum state. for a similar story for Wilson loops see \cite{Drukker:2006zk, Gomis:2008qa, Halmagyi:2007rw}. In simple terms they describe wavefunctions of semi-classical BPS states in $\mathcal{N}=4$ SYM and as such they provide a quantum mechanical description of half-BPS Coulomb branch configurations.  This makes them ideal candidates for computing quantities that one can match to the dual geometric description. In fact, we argue that despite the fact that one expects an exact match for half-BPS observables on both sides of the duality due to a lack of $g^2 N$ corrections to the free-field theory answer, the corresponding supergravity will not in general compute a precise quantity from the point of view of the conformal field theory but rather a (micro-canonical) average. This is purely an effect of the large $N$ limit and the full stringy description should be able to resolve the details of the boundary observable. These observations have been made in this context before, for instance in \cite{Skenderis:2007yb}, but we clarify how this happens on the field theory side of the computation. Our main result is a large $N$ formula for all heavy-heavy-light structure constants of half-BPS operators, for instance
\begin{equation}
C_{R\,R\,L/2}= \frac{1}{\sqrt{L}} \int dz d\bar{z} \, \rho_R(z,\bar{z}) \, (z+\bar{z})^{L},
\end{equation}
where the density $\rho_R$ is determined entirely from Young diagram data in a well-known way \cite{deMelloKoch:2008hen}. This is essentially the formula motivated in \cite{Skenderis:2007yb} from holographic renormalization of low lying operators and Coulomb branch limits. Our computation provides a check of this one-point function formula for all single trace primaries and in principle for all LLM geometries without relying on free-fermion methods. We also compute off-diagonal structure constants between sufficiently close heavy states suggesting that semi-classical supergravity calculations should be able to probe the precise microstructure of bubbling geometries.

This paper is structured as follows. In section 2, we review the BPS coherent state construction and and discuss the computation of the form factor of a single trace operator in the background of a giant graviton. In section 3 we generalize the computation to the case where the number of giant gravitons scales with $N$. To do this we explain how to reduce the corresponding integral over a Grassmannian to  a more conventional matrix model involving square matrices and  then solve the model at large $N$. The resulting distribution essentially reproduces the distributions studied in \cite{Chen:2007du}, up to a change of variables. We then study the general problem of multiple stacks of giant gravitons using steepest decent methods. In section 4 we return to the problem of computing correlators in the character basis and provide a more explicit connection from the eigenvalue picture presented by the coherent state generating functions and the character basis. In section 5 we conclude by discussing some general lessons and future directions.
\section{Coherent States and Form Factors}
\renewcommand{\theequation}{2.\arabic{equation}}
\setcounter{equation}{0}
Most of our discussion will concentrate on the simplest correlation functions in the $\mathcal{N}=4$ SYM theory, which are three point functions of half-BPS operators. We will also work mostly with the theory on the cylinder $\mathbb{R}\times S^3$, but translating the results to the plane is straightforward. A convenient parametrization for half-BPS operators is given in terms of a six dimensional null complex vector $n\cdot n=0$:
\begin{equation}
    Z(x, n)= n_I \phi^I(x),
\end{equation}
and any half-BPS operator is obtained by taking gauge invariant combinations of $Z(x,n)$. One common choice of operators are single and multi-trace operators
\begin{equation}
    \mathcal{O}_{\left\{L_i\right\}}(x)= \prod_{k}\Tr_N\left[\left(\vec{n} \cdot \vec{\phi}(x)\right)^{L_k} \right].
\end{equation}
In the large $N$ limit with $\Delta= \sum_{k} L_k\ll \sqrt{N}$ these provide an approximately orthogonal basis of operators; this is the usual statement that in the large $N$ limit planar graphs contribute the most in correlation functions. This class of operators is naturally associated to the supergravity modes of $AdS_5\times S^5$ and their bound states. However, for operators with large enough conformal dimension $\Delta$, various non-planar effects can contribute meaningfully or eventually dominate over planar graphs. Even more strikingly, certain extremal correlators of single traces operators have enhanced contributions from non-planar diagrams even for small charges \cite{DHoker:1999jke}. For these reasons it is useful to first perform the computation at finite $N$ with a proper orthogonal basis of states, and then take the large $N$ limit.

By restricting to primary operators, we will often drop the space coordinates $x_{1,2,3}$ since we will mostly work with constant modes on the $S^3$. Due to non-renormalization properties of half-BPS operators, the two and three point functions of such operators can be computed in the free field theory limit $g_{YM}=0$, so that our task reduces to a combinatorial problem of performing Wick contractions of free fields. This problem was first addressed in \cite{Corley:2001zk} for extremal correlators $n_1=n_2=n_3^*$. The main idea is to construct an orthogonal basis of states for one matrix quantum mechanics with $U(N)$ gauge symmetry, or equivalently a set of operators that diagonalize two point functions. The resulting basis is build from characters of the unitary group and is often referred to as the Schur basis:
\begin{equation}
    \mathcal{O}_R\left( Z(x, n)\right)= \frac{1}{k!}\sum_{\pi \in S_k} \chi^R(\pi) \Tr_{\mathbf{N}^{\otimes k}}\left[ \pi Z(x, n)^{\otimes k} \right]= s_R\left( Z(x, n)\right),
\end{equation}
where $k$ denotes the number of boxes of the Young diagram associated to the representation $R$ of $U(N)$ and $\chi^R(\pi)$ is the character of the corresponding representation of $S_k$. A different proof that this set of operators provides a diagonal basis was given in \cite{Berenstein:2022srd}. These operators have a dual description in terms of giant gravitons, or their bound states, in asymptotically $AdS_5\times S^5$ spaces \cite{Corley:2001zk, berenstein2004toy, Hashimoto:2000zp}.

Instead of performing the explicit contractions for a particular operator, it was realized that one could instead work with a coherent state for the free field $ Z(x, n)$
\begin{equation}\label{coherent state}
    \ket{\Lambda}= \frac{1}{\texttt{Vol}\left[U(N)\right]}\int d U \; e^{\Tr_N\left[ U \Lambda U^\dagger  Z(x, n)\right]} \ket{0}.
\end{equation}
A straightforward computation gives an alternate formula for $\ket{\Lambda}$ as a expansion in characters of the unitary group $s_R$,
\begin{equation}
    \ket{\Lambda}= \sum_{R} \frac{1}{f_R} \mathcal{O}_R\left( Z(x, n)\right) s_R(\Lambda),
\end{equation}
and $s_R(\Lambda)$ is a Schur polynomial. The point of this analysis is that by exploiting the Campbell-Hausdorff formula the free field contractions of the operators $  \mathcal{O}_R\left( Z(x, n)\right)$ can all be replaced by an integral over the unitary group.  For the two point functions the resulting integral is a Harish-Chandra-Itzykson-Zuber integral which has an exact fixed point formula:
\begin{equation}\label{HCIZ}
    \bra{\bar{\Lambda}} \ket{\Lambda}= \frac{1}{\texttt{Vol}\left[U(N)\right]}\int d U \; e^{\Tr[U\Lambda U^\dagger \Bar{\Lambda}]}= \mathcal{C}_N \sum_{\pi \in S_N} \; \det(\pi)\,\frac{e^{\lambda_i \Bar{\lambda}_{\pi(i)}}}{\Delta(\Lambda)\Delta(\Bar{\Lambda})}.
\end{equation}
Following the ideas of \cite{Jiang:2019xdz, Chen:2019gsb, Budzik:2021fyh}, one can reduce the computation of any correlator in the free theory to a matrix integral by commuting various generating functions past each other using the Campbell-Hausdorff formula. In the language of \cite{Jiang:2019xdz}, this is equivalent to replacing the fields inside small operators (such as traces) by their vevs after integrating out the SYM fields and then performing a saddle point approximation over the auxiliary parameters (in this case $U$). 

More concretely we will be interested in computing form factors such as:
\begin{equation}
\bra{\Bar{\Lambda}, n_3 } \Tr[Z(t=0, n_2)^L] \ket{\Lambda, n_1}\simeq \sum_{R, R'} C_{R,R'}(\Lambda, \bar{\Lambda}) \bra{R', n_3 } \Tr[Z(t=0, n_2)^L] \ket{R, n_1},
\end{equation}
where the initial and final states are created by heavy operators $\Delta \sim N^2$ in the large $N$ limit. 
For relatively simple choices of operators, such as determinants and traces of fully symmetric tensors \cite{Jiang:2019xdz, Chen:2019gsb, Yang:2021kot, Holguin:2022zii}, the saddle point analysis can be performed rather explicitly and the correlators can be matched precisely to their holographic counterparts. For more complicated operators, such as insertions of many determinant operators, or operators associated to generic Young diagrams, the saddle point analysis appears to be less straightforward and the structure of the solutions to the saddle point equations is not fully understood. The main difficulty lies in the fact that the resulting matrix models cannot be easily reduced to integrals over eigenvalues, so that the saddle point equations appear to be truly matrix equations. We will discuss in the later sections how to overcome these complications in the regimes relevant to states with nice supergravity descriptions (i.e. states corresponding to non-trivial geometries with small curvatures). 

\subsection{Example: AdS giant graviton}
Before proceeding to the case of interest, it is convenient to review the results presented in \cite{Jiang:2019xdz, Holguin:2022zii} since many of the parts of the calculations presented there extend naturally. In the simplest of cases, the heavy operators can be taken to be of rank one, meaning that they correspond to Schur polynomials of fully symmetric or fully anti-symmetric representations. In the case of fully anti-symmetric representations the correct generating function is the determinant operator \cite{Berenstein:2013md}, for instance :
\begin{equation}
    \det(\phi_5+i\phi_6-\lambda)= \det(Z-\lambda),
\end{equation}
which describes a sphere giant graviton sitting at the origin of global AdS at a position inside $S^5$ given by $e^{i \phi_0}\cos\theta_0= \lambda$. This operator is a semi-coherent superposition of all subdeterminant operators, each describing the R-charge eigenfunctions of a sphere giant graviton. The method for semi-classical computation with this class of operators was presented in \cite{Jiang:2019xdz} and also \cite{Budzik:2021fyh} so we refer the reader there for details. Instead we will describe how the analogous computation is done for an operator describing a semi-classical AdS giant graviton. The reason for this is that in the end both calculations lead to very similar integrals for the correlators, but their form is much easier to understand for the computation involving symmetric tensors. 

First we consider the following coherent state:
\begin{equation}
    \ket{\lambda}= \frac{1}{\texttt{Vol}\left[\mathbb{CP}^{N-1}\right]}\int_{\mathbb{CP}^{N-1}} d \varphi d\varphi^\dagger \; e^{\lambda \varphi^\dagger Z \varphi} \ket{0}.
\end{equation}
As discussed in \cite{Berenstein:2022srd, Holguin:2022zii} this is the same state that one obtains from setting $\Lambda$ to be a rank one projector in \eqref{HCIZ}. This state has a natural $U(1)$ gauge symmetry 
\begin{equation}
    \varphi\sim e^{i\alpha} \varphi,
\end{equation}
which can be identified with the gauge symmetry on the worldvolume of the giant~graviton, as well as invariance under $U(N)$ gauge transformations of $Z$. This state is also a coherent superposition of AdS giant graviton wavefunctions with fixed R-charge. A simple calculation yields:
\begin{equation}
\bra{\bar{\lambda}}\ket{\lambda}= \frac{1}{\texttt{Vol}\left[\mathbb{CP}^{N-1}\right]}\int_{\mathbb{CP}^{N-1}} d \varphi d\varphi^\dagger \; e^{\lambda \bar{\lambda} \varphi^\dagger P_1 \varphi}.
\end{equation}
To evaluate this integral we need to do a series of simple coordinate transformations. Without loss of generality, we can let $P_1$ be a rank one projector into the first component of $\varphi$. Then we can split the coordinates of $\mathbb{CP}^{N-1}$ into $\varphi_1$ and $\varphi_{n}$ with $n>1$. The reason we emphasize this will become clear when we generalize this more complicated coherent states. Then, we can parametrize the coordinates $\varphi_n$ in terms of an $2N-2$ dimensional spherical slices of radii $R=\sqrt{1-|\varphi_1|^2}=\sqrt{1-r^2}$.
Finally we can rewrite the radial part of the integral as 
\begin{equation}
   \int d\left(r^2\right) d\left(R^2\right) \; R^{2N-4}\; \delta(R^2+r^2-1) \; e^{\lambda \bar{\lambda} r^2}=  \int_0^1 dx \; (1-x)^{N-2} \; e^{\lambda \bar{\lambda} x}.
\end{equation}
This last integral is simply the moment generating function of a particular unitary Jacobi distribution. To make contact with the calculation involving determinants, we can rewrite 
\begin{equation}
    \left(\lambda \bar{\lambda} \right)^{N-2}(1-x)^{N-2}= \det\begin{pmatrix} \lambda&& \bar{\lambda}
\varphi_1 \\ \lambda \varphi_1^*&&\bar{\lambda} \end{pmatrix}^{N-2} = \det\begin{pmatrix} \rho_{11}&&\rho_{12}\\ \rho_{21}&&\rho_{22} \end{pmatrix}^{N-2}.
\end{equation}
Although this step is not necessary and the previous integral expression is simple enough to evaluate explicitly, doing this change of variables makes it clear that the final answer the large $N$ approximation for AdS giants is the same as that of sphere giants (up to analytic continuation). After a final simple re-scaling, $\lambda\rightarrow\sqrt{N-2} \;\lambda$ we finally arrive at the expression
\begin{equation}
    \bra{\bar{\lambda}}\ket{\lambda}= C_N \int_{\det \rho\;\geq \lambda \bar{\lambda}-2 } d\rho \; \det  \rho^{N-2}\; e^{\left(N-2\right)\tr_{2}\left[ \rho \rho^\dagger \right]},
\end{equation}
which apart from the contour on integration is identical to the integral obtained from the Hubbard-Stratonovich trick used in \cite{Yang:2021kot} for determinant operators \footnote{Our trace convention is so that $\tr_m 1=1$ as opposed to $\tr_m 1=m$ }. This will be true generically, once we solve the saddle point equations for a configuration of AdS giant gravitons, we automatically have the solution for a configuration of sphere giant gravitons after a simple analytic continuation. In this case the saddle point equations are simply:
\begin{equation}
\begin{aligned}
    \rho^\dagger&= \rho^{-1} \Rightarrow \rho_{12}\rho_{21}= \lambda \bar{\lambda}-1\\
    \lambda \bar{\lambda}&>1.
\end{aligned}
\end{equation}

The second equation comes from the fact that the exponential needs to be positive for the saddle point to be a maximum; this implies that the semiclassical approximation is valid whenever $|\lambda|= \cosh\rho_0$ is greater than one which simply says that the brane is at a position $\rho_0>0$ in global AdS, and this is true for all half-BPS giant graviton solution.
\subsubsection{Form Factors}
The next step is compute the following form factor;
\begin{equation}
    \bra{\bar{\lambda}} \Tr_N\left[\left( \vec{n}\cdot \vec{\phi}\right)^L\right] \ket{\lambda}=  \bra{\bar{\lambda}} \Tr_N\left[\left(\frac{ Z+\bar{Z}+Y-\bar{Y} }{2}\right)^L\right] \ket{\lambda}.
\end{equation}
Our choice of $\Vec{n}$ is taken from \cite{Yang:2021kot} for clarity of presentation and is arbitrary. To evaluate this quantity, we use the fact that the initial and final states are coherent states which lets us replace $Z$ and $\bar{Z}$ by constant matrices. The resulting trace is
\begin{equation}
   \Tr_N \left[\left(\frac{ Z+\bar{Z}+Y-\bar{Y} }{2}\right)^L\right]= \Tr_N\left[\left(\frac{ \bar{\lambda} \varphi\varphi^\dagger+ \lambda \tilde{\varphi} \Tilde{\varphi}^\dagger }{2}\right)^L \right].
\end{equation}
To proceed we use the fact that the integrals over $d \varphi$ and $d \tilde{\varphi}$ are invariant under the action of $U(N)$, so we can gauge fix $\tilde{\varphi}$ to be a unit vector $v$ with a one in the first component. Finally one uses the trick introduced in \cite{Jiang:2019xdz} to exchange the trace over color indices into a trace over ``flavor'' indices associated to the in and out states of the brane:
\begin{equation}
   \Tr_N \left[\left(\frac{ Z+\bar{Z}+Y-\bar{Y} }{2}\right)^L\right]= 2^{-L}\tr_2\left[\begin{pmatrix}  \lambda&& \bar{\lambda}
\varphi_1 \\ \lambda \varphi_1^*&&\bar{\lambda}\end{pmatrix}^{L}\right]\simeq \tr_2 \rho^L.
\end{equation}
Now instead of evaluating  every power the matrix $\rho$, it is better to work with the resolvent
\begin{equation}
    R(t)= \tr_2\left[\left( 1- t\rho \right)^{-1} \right]= \frac{1- t \tr_2[\rho] }{t^2\det\rho-2 t \tr_2[\rho]+1 }.
\end{equation}

When we evaluate this expression at the saddle point value for $\rho$, we can immediately recognize that the resolvent $R(t)$ is a generating function for Chebyshev polynomials of the first kind.
\begin{equation}
    \left< R(t)\right>_{N=\infty}= \sum_{n=0}^\infty T_n( \cos \phi_0 \cosh \rho_0)\;  t^n,
\end{equation}
where we parametrized the eigenvalue in terms of LLM coordinates $\lambda= e^{i \phi_0} \cosh\rho_0$.
\subsubsection{Extracting Structure Constants}
Naively one might expect that the saddle point approximation of the resolvent computes a generating function of some half BPS structure constants. This is not quite correct for the following reason. First we would need to extract the contribution to the form factor from a particular set of primary operators. In this case this is somewhat easy to do, given that the coherent state has a simple expansion in terms of Schur polynomials for rank one representations
\begin{equation}
    \ket{\lambda}= n_\lambda \sum_{k=0}^\infty\frac{\lambda^{N+k-1}}{(N+k-1)!} \mathcal{O}_{(k)}(Z) \ket{0}.
\end{equation}
We should then think of the coefficients in the expansion as the distribution of lengths for a Young diagram with a single row. The distribution is similar to a Poisson random variable, so that the average length is of the Young diagram is of order $|\lambda|$. Another way of seeing this is by using the Stirling formula for the denominator and extremizing with respect to $l=N+k-1$; the maximum occurs when $|\lambda|= N+k-1$. This is the reason why the coherent state calculation in \cite{Holguin:2022zii} gives the correct answer for structure constants without the need to project into a particular character. 
However, the dependence on the phase of $\lambda$ when we insert an operator will not be correct due to unwanted contributions coming from off-diagonal terms. To fix this one should project the intermediate operator into a an $R$-charge singlet operator. This is done by performing a group average over the phase of $\lambda$. Since the wavefunction $\bra{\bar{\lambda}}\ket{\lambda}$ is already invariant under shifts in the phase of $\lambda$ whenever $\bar{\lambda}=\lambda^*$, the only effect of averaging is to project out off-diagonal terms from the resolvent. 

Strictly speaking this averaging should be performed prior to doing the saddle analysis, since averages do not generally commute. One way of performing this average is to rescale $t\rightarrow \sqrt{\det \rho} \;t$, and then perform the integration over the phase of $\lambda$ by a contour integral. The result is 
\begin{equation}
    \overline{R(t)}= \sqrt{\det \rho}\left[\frac{t^2-1}{2\sqrt{(1+t^2)^2\det\rho- 4t^2 \lambda \Bar{\lambda}}}\right].
\end{equation}
In the large $N$ limit, we can set $\det \rho=1$, and the averaged resolvent will take the form of a generating function of Legendre polynomials. Since the saddle point analysis for this case basically involves setting $\rho$ to a particular value, the averaging procedures commute so the large $N$ limit was taken first in \cite{Yang:2021kot, Holguin:2022zii} without any trouble. This is not the case for operators made out Schur polynomials for large Young diagrams, since the large $N$ limit leads to a continuous distributions of eigenvalues. Once the eigenvalues condense, the result of the computation will be highly sensitive to the analytic properties of the moment generating function, and performing the averaging and large $N$ limit can lead to contradicting results. The most natural prescription to remedy this is to perform the projection into a particular primary operator first by an appropriate averaging, and then take the large $N$ limit of this quantity.
\section{Matrix models for general coherent states}
\subsection{Two Droplets}
\renewcommand{\theequation}{3.\arabic{equation}}
\setcounter{equation}{0}
The next simplest calculation that we can perform is the case where the matrix $\Lambda$ in \eqref{HCIZ} is taken to be a rank $p$ projector. In this case, the analogous coherent state is an integral over the Grassmannian $Gr(p, N)$. The expression for the coherent state is simple to write down, but some of the steps needed to evaluate the resulting matrix integrals require some care; the main task will be to evaluate the following norm:
\begin{equation}
    \ket{\lambda, p}= \frac{1}{\texttt{Vol}\left[Gr(p, N) \right]} \int_{Gr(p, N)} dV dV^\dagger \; e^{\lambda\Tr_N{V V^\dagger Z}}\ket{0}.
\end{equation}
We will argue that this state described the wavefunction of a stack of $p$ giant gravitons sitting at position $\lambda$ in the LLM plane. The first thing to note is that this coherent state has an explicit $U(p)$ gauge symmetry $V\sim V g$ which we can identify as the gauge symmetry on a stack of D-branes. The expectation value of $\Delta_0$ on this state is given by $p |\lambda|^2$ so that whenever $|\lambda|\sim \sqrt{N}$ and $p\sim N$, the average dimension of this state is of order $N^2$. By inspection, we can also deduce that this state is a coherent superposition of Schur operators of at most $p$ rows. by acting with $\Tr[\bar{Z}]$ on this state, we can see that this state breaks the gauge symmetry spontaneously from $U(N)$ to $U(N-p)\times U(p)$, and that the center of mass of the stack of $p$ branes is at the position $z=\lambda$ on a complex plane.

Before proceeding we need to comment on the choice of coordinates for Grassmannian, since the details on how to perform these types of integrals are known but not readily available. We will mostly follow the notation of \cite{adler2001integrals, duenez2003random}; for a pedagogical presentation we refer the reader to \cite{Livan_2018}. First, we can choose to split any given group element $U$ in terms of block matrices

\begin{equation}
    U = \begin{pmatrix} A_{11}&& A_{12}\\ A_{21}&& A_{22}\end{pmatrix},
\end{equation}
where $A_{11}$ and $A_{22}$ are $p\times p$ and $(N-p)\times(N-p)$ square matrices, and $A_{12}$ is a $p\times (N-p)$ matrix (similarly for $A_{21}$). Then we make an arbitrary choice of frame distinguished by a rectangular matrix 
\begin{equation}
    v^T= \begin{pmatrix}
      \;  \mathbb{I}_p && \boldsymbol{0}_{(N-p)\times p};
    \end{pmatrix}
\end{equation}
this matrix can then be used to build projectors into arbitrary $p$-dimensional subspaces of $\mathbb{C}^N$ by acting on $v$ with unitaries. This set of projectors precisely gives a parametrization of the affine Grassmannian  $Gr(p, N)$. By an affine Grassmanian we will simply mean the space spanned by the unitary transformations of $v$:
\begin{equation}
    Gr(p,N):= \{ V = U \cdot v \; | \; U \in G \}.
\end{equation}
Clearly any $V$ in this space is rank deficient, so all of the information about its singular value decomposition is captured a square matrix:
\begin{equation}
    V V^\dagger= U^\dagger P_p U= \begin{pmatrix}A_{11}^\dagger A_{11}&&0\\0&&0
        
    \end{pmatrix} .
\end{equation}
In analogy to the rank one calculation, we can write down the integration measure for this space as
\begin{equation}
    dV dV^\dagger= dA_{11}dA_{11}^\dagger dA_{12}dA_{12}^\dagger \times \delta\left(A^\dagger_{11} A_{11}+ A_{12}^\dagger A_{12}-\mathbb{I}_p \right)
\end{equation}
 By a similar calculation we can see that the norm of this state is given by 
\begin{equation}
    \bra{\bar{\lambda},p\;}\ket{\;\lambda, p}= \frac{1}{\texttt{Vol}\left[Gr(p, N) \right]} \int_{Gr(p, N)} dV dV^\dagger \; e^{\lambda \bar{\lambda}\Tr_p{ V^\dagger P_p V}}.
\end{equation}
One last fact that we need before continuing is the Jacobian for the coordinate transformation $M= A^\dagger A$ for any $n\times m$ rectangular matrix $A$. This change of variables is well known in the context of Wishart distributions;
\begin{equation}
    dAdA^\dagger \propto \det(M)^{n-m}dM,
\end{equation}
where the constant of proportionality is an integral over angular variables that will not be important for our analysis. So in the end we find 
\begin{equation}\label{normp}
      \bra{\bar{\lambda},p\;}\ket{\;\lambda, p}= n_{\lambda, p} \int_{A_{11}^\dagger A_{11}\preceq\; \mathbb{I}_p} dA_{11}^\dagger dA_{11} \det\left(\mathbb{I}_p-A_{11}^\dagger A_{11} \right)^{N-2p} \exp\left( \lambda \bar\lambda\Tr_{p}\left[ A_{11}^\dagger A_{11} \right]\right).
\end{equation}
Since the only combination of $A_{11}$ and $A_{11}^\dagger$ that appears in the integral is $A_{11}^\dagger A_{11}$, we can reduce the computation to an integral over the eigenvalues of $A_{11}^\dagger A_{11}$.
\begin{equation}\label{partition function}
    \bra{\bar{\lambda},p\;}\ket{\;\lambda, p}= \mathcal{C}_{\lambda, p} \int_{[0,1]^p}\prod_{i=1}^p dx_i\; \Delta\left(x_i\right)^2 \left(1-x_i\right)^{N-2p}\; \exp\left( \lambda \bar{\lambda}\sum_{j=1}^p x_j\right).
\end{equation}
\subsection{Large $N$ Limit: Steepest Descent}
We will now sketch the evaluation of \eqref{partition function} in the large $N$ limit. As before, $|\lambda|^2$ will scale with $N$ so that the exponential term is large. The integral can evaluated explicitly using the Andreief identity \cite{Livan_2018}:
\begin{equation}
    \mathcal{C}_{\lambda, p}^{-1}= p! \det_{j,k}\left(\int_{0}^1 dx  \;(1-x)^{N-2p} x^{j+k-2} e^{\lambda \Bar{\lambda} x}\right)
\end{equation}
The function inside the determinant is an incomplete Gamma function. Another way interpreting it is at the moment generating function for the GUE Jacobi ensemble. Even though we can evaluate simple moments in this distribution exactly, the final form will always be a large determinant of a Hankel matrix which we cannot deal with easily. Instead we can perform a saddle point analysis of \eqref{partition function} for large $N$ and large $p$. After rescaling a similar rescaling $\lambda \rightarrow\sqrt{N-2p}\;\lambda$ the saddle point equations of this integral are of the form
\begin{equation}
    \lambda \bar{\lambda}- \frac{1}{1-x_i}- \frac{2}{N-2p}\sum_{j\neq i}\frac{1}{x_j-x_i}=0.
    \end{equation}

The behavior of the saddle point configurations are easy to understand; the first two terms are the same as in the single eigenvalue problem so that all the eigenvalues have a tendency to condense around $\lambda\bar{\lambda}=\frac{1}{1-x_i}$ while second term the usual eigenvalue repulsion term. We should also note that for distribution to be stable we have to require that $p\leq N/2$. This makes sense, since condition forces the second droplet to be smaller than the first. In the case that $p\gg N/2$, this configuration is no longer a small deformation of the original vacuum configuration and the remaining $N-2p$ eigenvalues become the relevant degrees of freedom. The case with $p\sim N/2$ has to be treated with particular care as we will see.  To solve these equations at large $N$ with $p/N$ fixed we simply recast the saddle point equations as a Ricatti equation for the resolvent matrix $R_p(z)$:
\begin{equation}\label{one-ev-res}
    \left[\lambda \bar\lambda- \frac{1}{1-z} \right] R_p(z)- \frac{K}{1-z}- \frac{p}{N-2p} R_p(z)^2+ \frac{1}{N-2p} R'_p(z)=0,
\end{equation}
where $K=\frac{1}{p}\sum_{i=1}^p \frac{1}{1-x_i}= \int \frac{\rho(z)}{1-z}$ is a constant that is determined from $R_p(z)$ by imposing self-consistency conditions of the distribution. From this we can identify $\frac{1}{N-2p}$ as the relevant $\hbar$ parameter in the problem; this is the reason why $p\sim N/2$ should be treated with care, since the $1/N$ fluctuations of the resolvent are no longer under control and one must solve the differential equation exactly. This can be seen from a simple scaling argument; if $N-2p$ is of order one, the exponential and determinant terms in \eqref{normp} are of order $e^p$, while the Vandermonde term is of order $e^{p^2}$. This says that the dominant effect in this case is the eigenvalue repulsion, so that the separation of each of the individual eigenvalues is large when compared to the size of the system. In particular we should expect the corresponding geometry to have a region with string scale curvature where the supergravity approximation breaks down. This is expected, since wavefunction $\ket{\lambda,p}$ no longer has a good semiclassical approximation in the large $N$ limit. If instead we decide to keep $N-2p$ of order one but this time scaling $\lambda$ as $\sqrt{p}\sim \sqrt{N}$, then the Vandermonde contribution is off-set by the exponential term and the determinant term is still sub-leading, so  the distribution of eigenvalues $x_i$ is approximately semi-circular. What this is saying is that now the two droplets are too close together to be treated as separate (meaning that their size is of the same order as their separation), and instead the deviation from vacuum is described by a collection of order one giant gravitons probing the vacuum.  In other words, depending on how we decide to scale $N-2p$ we will obtain qualitatively different saddle point conditions and the expansion in terms of AdS or sphere giant gravitons might be more suitable.

The regime we will be interested in is when $N-2p$ is of order $N$, so that the constant equilibrium configuration is a good approximation to the eigenvalue distribution. In the large $N$ limit the density of states becomes
\begin{equation}\label{distribution}
    \rho(x)dx=\frac{1}{\pi}\frac{\sqrt{4 K \mu(1-x)-\left( 1-(1-x)\lambda \bar{\lambda}\right)^2}}{1-x} dx,
\end{equation}
with $\mu= \frac{p}{N-2p}$; this turns out to be simpler after we make the change of variables $z=\lambda\bar{\lambda}\left(1-x \right)$. After this we can normalize the distribution such that $\int \rho(z) dz= 2\mu$  and find $K= 1$. In this convention the eigenvalues are quantized in units of $2 g_s= \frac{1}{N-2p}$, where $g_s$ is the effective string coupling of this system. 

To compare with the corresponding LLM solution we can express the solution in coordinates that manifest $\frac{1}{8}$ of the supersymmetries \cite{Chen:2007du}. The point is to write the 10d metric in terms of a 6d complex basis with coordinates $x,y,z$. For the vacuum $AdS_5\times S^5$ solution these coordinates should be identified with the coordinates of the five-sphere. Translating the whole metric into these coordinates is a non-trivial task for generic LLM geometries, but we will only be interested in determining the volume of the cycle wrapped by the branes. In these coordinates the radius of the three-sphere wrapped by the giant gravitons for a single droplet solution is given 
\begin{equation}
    \tilde{r}= \frac{\sqrt{(L^2-|z|^2)(|z-a|^2-b^2)}}{|z-a|},
\end{equation}
where the droplet is centered at $z=a$, the size of AdS is $L$ and $b$ is the radius of the droplet. Notice that this is essentially of the same form as \eqref{distribution} up to some relabellings. The discrepancy between the denominators is due to the fact that the variable $x$ is actually related to the square of the radial direction of AdS. The precise mapping between both pictures should involve some more complicated charge of variables in general, but the analytic properties of both distributions are the same.
\subsection{Three Point Functions: Diagonal Case}
To compute the correlator of a single trace in the background of these coherent states we can use the resolvent trick. Using the same kind of color-flavor transformation the trace over the original color indices can be replaced by a trace over $p\times p$ matrix. In this case the moment generating function is 
\begin{equation}
\begin{aligned}
    \mathcal{F}(t)&= \tr_p\left[\left((2-t(\lambda +\bar{\lambda}) \right)\left( (1-t \lambda)(1-t\bar{\lambda})\mathbb{I}_p- t^2 \lambda A_{11}^\dagger A_{11}\right)^{-1}\right]\\
    &= 2\sum_{i=1}^p \frac{1- \frac{t}{2}(\lambda+\bar{\lambda}) }{t^2 \lambda \bar{\lambda}(1-x_i) - t(\lambda +\bar{\lambda})+1}.
\end{aligned}
\end{equation}
The most natural variable to work with is once again $z_i=\lambda\bar{\lambda}(1-x_i)$, which makes the density of eigenvalues be of the form:
\begin{equation}\label{marchenko-pastur}
    \rho(z)= \frac{\sqrt{(z_+-z)(z-z_-)}}{\pi z},
\end{equation}
where $z_\pm $ are given by the roots of the polynomial inside the square root in \eqref{distribution}
\begin{equation}
    z_\pm= 1+\frac{2 \mu}{\lambda\bar{\lambda}}\pm\frac{2\sqrt{\mu(\mu+\lambda\bar{\lambda})}}{\lambda\bar{\lambda}}.
\end{equation}
Expanding $\mathcal{F}(t)$ as a function of $z$ gives an expression for the moment generating function in terms of the moments of the Marchenko-Pastur distribution \eqref{marchenko-pastur}. After extracting the $L^{\text{th}}$ moment we get
\begin{equation}
\begin{aligned}
 &\bra{\bar{\lambda}} \Tr[\frac{(Z+\bar{Z}+Y-\bar{Y})^L}{2^L}]\ket{\lambda}=\\ &\sum_{k=0}^{\infty} \binom{\frac{L+k}{2}}{\frac{L-k}{2}} 2^{-(L-k)}\left(\frac{\lambda+\bar{\lambda}}{2}\right)^k\, m_{\frac{L-k}{2}}
 - \sum_{k=0}^{\infty} 2^{-(L-k)}\binom{\frac{L+k}{2}-1}{\frac{L-k}{2}} \left(\frac{\lambda+\bar{\lambda}}{2}\right)^{k+1}\, m_{\frac{L-k-1}{2}},\\
 &m_l= \int_{z_1}^{z_+}dz \, \rho(z)\, z^l= \sum_{k=1}^l \frac{1}{k}\binom{l}{k}\binom{l}{k-1} \left( \frac{z_+-z_-}{2}\right)^k=\;_2F_1\left(1-l,-l;\, 2;\
 , \frac{z_+-z_-}{2}\right)
\end{aligned}
\end{equation}
To extract the diagonal part of this form factor we can average over the phase of $\lambda$. 
\begin{equation}
\mathcal{F}_{\lambda\, \bar{\lambda}\,L}= \frac{1^L+(-1)^L}{2 \sqrt{L}} \sum_{k=0}^{L} \binom{\frac{L+k}{2}}{\frac{L-k}{2}} \binom{k}{\frac{k}{2}} \left(\frac{1^{k}+(-1)^k}{2}\right)\, |\lambda|^k\,\; m_{\frac{L-k}{2}} \, \left(\frac{L}{L+k}\right).
\end{equation}
This correlator should be interpreted as encoding part of the angular distribution of the bubbling geometry associated to the condensate of eigenvalues and should compute the one-point function of a scalar operator on an LLM geometry with two circular droplets. It would be interesting to compute these correlators holographically, for instance with the methods developed in \cite{Skenderis:2007yb}. By performing an additional contour integral over $|\lambda|$ we can obtain three point functions for operators with fixed scaling dimensions as opposed to coherent states as is done in \cite{Yang:2021kot}. One can similarly perform computations that extract off-diagonal form-factors between heavy states with different dimensions. 

\subsection{General Matrix Model: Eigenvalue Picture}
\renewcommand{\theequation}{1.\arabic{equation}}
\setcounter{equation}{0}
Now we will proceed to the general case and sketch how to extract specific operators associated to a particular Young diagram of at most $p$ rows with order $N^2$ boxes, and we will use this to compute diagonal form factors. To do this we need to consider the generating function
\begin{equation}
    \bra{\Bar{\Lambda}_p} \ket{\Lambda_p}= n_{p} \int d\sigma^\dagger d\sigma \det\left( \mathbb{I}-\sigma^\dagger \sigma\right) \; e^{N\tr_p[\Lambda \sigma \Bar{\Lambda}\sigma^\dagger]}.
\end{equation}
This formula is the analog of \eqref{normp} for generic coherent state parameters. 
Computing this integral is not an easy task for generic eigenvalues, since the argument of the exponential is no longer just a function of $\sigma^\dagger \sigma$, and the matrix $\sigma$ is not a normal matrix, so $\sigma$ and $\sigma^\dagger$ cannot be simultaneously diagonalized. In this section we will give a heuristic argument for a saddle point approximation to this integral. We will give a more concrete proof in the next section where we will need a more careful analysis of these type of integrals. Intuitively one would like to say that the integral is dominated by points where the all the matrices in the trace are diagonal. One way to see why this could be true is that the exponent can be written in the form:
\begin{equation}
    \tr_p\left[\Lambda \Bar{\Lambda} \sigma \sigma^\dagger - \sigma \Bar{\Lambda} [\Lambda,\sigma^\dagger] -\Lambda \Bar{\Lambda} [\sigma, \sigma^\dagger]  \right].
\end{equation}
Since the first term is manifestly positive, the exponent is the largest when this term is maximized which happens whenever all the matrix components are concentrated on the diagonals. Any deviation from this contributes to the second and third terms, which are not necessarily positive. So it is natural to expect that a good approximation to the integral is obtained by integrating over the set of $\sigma$ satisfying
\begin{equation}
    [\lambda, \sigma^\dagger]=[\sigma, \sigma^\dagger]=0.
\end{equation}
Indeed we will see later that these are the saddle point conditions for the integration over the angular variables for $\sigma$. Because of the large exponent, corrections to this are heavily suppressed as long as $\Lambda^\dagger=\bar{\Lambda}$, and so
\begin{equation}
     \bra{\Bar{\Lambda}_p} \ket{\Lambda_p}\simeq \int_{[0,1]^p}  dx_i \Delta_p(x_i)^2\, (1-x_i)^{N-2p}\, e^{N\sum_i |\lambda_i|^2 x_i} + O(e^{-N^2}).
\end{equation}
The other saddle points contribution for the norms consists of pairing the eigenvalues $\lambda_i$ with $\bar{\lambda}_{\pi(i)]}$ for all permutations $\pi$, which are highly suppressed for non-coincident eigenvalues. 
\begin{equation}
    e^{\Tr[\Lambda \sigma \Bar{\Lambda}\sigma^\dagger]}\rightarrow\sum_{\pi\in S_p} e^{\sum_{i } |s_i|^2 \lambda_i\bar{\lambda}_{\pi(i)}}\times\left( \texttt{one-loop}\right).
\end{equation}
We work out the appropriate one-loop determinant for this saddle point approximation in section 4. This structure precisely explains the saddle point structure found in \cite{Jiang:2019xdz} for determinant operators; the solutions to the matrix form of the saddle point equations always involve summing over permutations of initial and final giant graviton states.
For widely separated eigenvalues the Vandermonde determinant does not contribute meaningfully to the saddle point approximation and the state is well approximated by a collection of widely separated giant gravitons. A more interesting regime is whenever we have $n_a$ coincident branes at a point $\lambda_a$, for $a=1,\dots,k$. As long as the $\lambda_a$ are sufficiently separated interactions between different droplets can be neglected and the eigenvalues $x_i$ are distributed along $k$ cuts whose distribution is approximately given by the Marchenko-Pastur distribution. More precisely, the norm of the coherent state with $k$ lumps of eigenvalues centered around $\lambda_a$ is computed by the following matrix model:
\begin{equation}
    \mathcal{Z}(\lambda_1, \dots, \lambda_k)= \int_{[0,1]^p} \prod_{a=1}^k \,\prod_{i_a}^{n_a} dx_{i_a}^{(a)}\, \Delta_{n_a}\left(x^{(a)}\right)^2\,\left(1-x_{i_a}^{(a)} \right)^{N-2p}  e^{N\lambda_a\Bar{\lambda}_a x_a}\times \prod_{c>b}\,\prod_{i_b, j_c} \left(x^{(c)}_{j_c}- x^{(b)}_{i_b}\right).
\end{equation}
The derivation of this class of models was presented in \cite{Halmagyi:2007rw}, and we outline the details in the next section. As long as the cuts are not exponentially close to one another, the last eigenvalue repulsion term is far enough from zero that it does not affect the saddle point. The analysis for the three point function is then relatively straightforward; the moment generating function can be block diagonalized and each block is dealt just like the single-cut case. In the regimes when two cuts approach each other this approximation is no longer valid and one has to solve the corresponding monodromy problem exactly as the fluctuations around the stationary eigenvalue distribution will not be suppressed. We expect that these corrections reproduce the supergravity picture of \cite{Chen:2007du}, with the eigenvalue distribution being related to the volume of the three-sphere on which the giant gravitons are wrapped (see for example equations (5.110) and (5.116) in \cite{Chen:2007du}). We expect that the spectral curve for the matrix model is precisely encoded by the dual LLM geometry written in the coordinates advocated by \cite{Chen:2007du}, since our distribution of eigenvalue for the single cut case is essentially identical to their equation (5.117). This aligns with the results coming from numerical tests performed in \cite{Berenstein:2007wz} and we advocate for a similar viewpoint; $1/N$ effects will generically give some amount of granularity to the edges of eigenvalue droplets, specially if multiple droplets are close to one another when compared to the characteristic size of each eigenvalue. In order to resolve these details one would need to solve the full interacting saddle point equations. We will not do this, since in this regime there will not be a reliable geometric picture for the state. In other words, the saddle point approximation we described above breaks down for states that describe half-BPS geometries with string scale curvature, and instead we should think of the state as being a deformation of a smooth geometry with some branes inserted. 
\subsection{Coulomb Branch Limit}
One last interesting limit that we can consider is the limit in which the droplets are widely separated from each other and from the origin. In this limit, the Marchenko-Pastur distribution reduces to a delta function
\begin{equation}
    \rho(x)\rightarrow \sum_i\delta\left(\lambda_i\bar{\lambda}_i -\frac{1}{1-x_i}\right).
\end{equation}This is exactly the Coulomb branch limit discussed in \cite{Skenderis:2007yb}. In this sense the operators we study here can be understood as quantum mechanical analogs of Coulomb branch vacua of the theory. This makes the relation between the geometry of the moduli space of vacua and asymptotically anti-de-sitter spaces clear; in the large $N$ limit, the moduli space should get quantum correction which deform its geometry into a bubbling geometry and only in the dilute gas approximation can we approximate such a geometry by a multi-center solution. 
\section{Matrix Models for the Character Basis}
\renewcommand{\theequation}{4.\arabic{equation}}
\setcounter{equation}{0}
Now we will concern ourselves with computing three point functions where the initial and final state are specified by specific Young diagrams, as opposed to a collection of eigenvalues $\lambda_i$. The idea will be to make a somewhat unconventional choice of integration contour for the coherent states parameters. First we start with a pair of states $\ket{\tilde{\Lambda}^\dagger},\,\ket{\Lambda}$, but now we treat the parameters as being independent from one another; we will also force each of the eigenvalues $\lambda_i$ and $\Tilde{\lambda}_i$ to lie on a unit circle. By multiplying $\ket{\Lambda}$ by the square of the Vandermonde determinant of $\Lambda$, and integrating we can recognize that the resulting integration measure is just a Haar measure for a new unitary matrix $\mathcal{U}= U \Lambda U^\dagger$:
\begin{equation}
    \int_{U(N)} dU \oint d\lambda_i \, \Delta(\lambda_j)^2 \rightarrow \int_{U(N)} d\, \mathcal{U}.
\end{equation}
The resulting state is clearly proportional to the vacuum state, since there are no $\mathcal{U}^\dagger$ insertions to feed to the exponential. From this it becomes clear that in order to extract a term proportional to the state $\ket{R}= S_R(Z)\ket{0}$ one should multiply the integrand by a character $S_R(\,\mathcal{U}^\dagger)$:
\begin{equation}\label{projector}
    \ket{R}=  \frac{f_R }{\texttt{Vol}\left[U(N)\right]}\int d\, \mathcal{U} \;e^{\Tr[\, \mathcal{U} Z]}\; S_R(\,\mathcal{U}^\dagger) \ket{0}.
\end{equation}
A similar trick was used in \cite{Okuda:2007kh} to study expectation values of Wilson loop for arbitrary representations in large $N$ Chern-Simons theory, with a slightly different generating function. For our choice of generating function the exponential factor can be expanded in terms of unitary characters using Schur-Weyl duality and the resulting integrals are easily evaluated using elementary orthogonality relations. On the other hand, for sufficiently large representations we will be able to perform the integral using steepest descent after we perform all contractions of the $\mathcal{N}=4$ SYM fields. This exponential generating function is also useful for computing Wick contractions with other operators since we can exploit the Campbell-Hausdorf formula. With all this in mind the coefficient $f_R$ is easily determined to be 
\begin{equation}
    f_R= n_R! \,\frac{\Tr_R[\, 1\,]}{\chi_R(\text{id})}= n_R!\,\frac{\text{Dim}_R(N)}{d_R}\,;
\end{equation}
which is the norm of the corresponding state; this is done by expanding the exponential and matching the terms as in \cite{Berenstein:2022srd}. 

We will want to compute quantities such as
\begin{equation}
   C_{R R' L}=2^{-L}\times \frac{\bra{R'}\Tr_N \left[\left( Z+\bar{Z}+Y-\bar{Y} \right)^L\right]\ket{R}}{\sqrt{L\bra{R'}\ket{R'} \bra{R}\ket{R}}},
\end{equation}
in the limit that $|R|\sim |R'| \sim N^2$. To compute the quantity in the numerator we can substitute the equation \eqref{projector} and perform the Wick contractions using the Campbell-Hausdorff formula:
\begin{equation}
\begin{aligned}
    \bra{R'}\Tr_N \left[\left( Z+\bar{Z}\right)^L\right]\ket{R}= \frac{f_{R'}f_{R}}{\texttt{Vol}\left[U(N)\right]^2} \int d\, \mathcal{U} d\,\mathcal{V} \;S_{R}(\mathcal{U}^\dagger)S_{R'}(\mathcal{V}^\dagger)\; e^{N\Tr[\mathcal{U}\,\mathcal{V}]} \Tr\left[\left(\mathcal{U}+\mathcal{V} \right)^L \right].
\end{aligned}
\end{equation}
This procedure replaces all of the free-field Wick contractions with unitary integrals which we can evaluate very explicitly.
\subsection{Diagonal Structure Constant}
As before it will be easier to work with the moment generating function for the matrix $U+V$ instead of dealing with each individual trace. We now proceed by diagonalizing both $U$ and $V$, after which we are left with an integral of HCIZ-type:
\begin{equation}
    \mathcal{F}(t)\simeq \int d \tilde{U} d\mu(u) d\mu(v) S_R(u^*) S_R(v^*) e^{N \Tr[\tilde{U}^\dagger u\,\tilde{U} v]} \Tr[\left( 1-t(u+\tilde{U}v\tilde{U}^\dagger) \right)^{-1}].
\end{equation}
This integral is quite challenging to evaluate exactly, mainly due to the appearance of the unitaries $U$ inside of the trace of the resolvent. At large $N$, the integral over $U$ can be evaluated the method of steepest descent; the saddle point equations for the matrix $U$ are solved by permutation matrices which allows us to replace the integral over $U$ by a sum over permutations times a one loop determinant
\begin{equation}
    \int d\tilde{U} \rightarrow\frac{1}{N!}\sum_{\pi\in S_N}\times \prod_j\frac{1}{\nu_i},
\end{equation}
where $\nu$ are the eigenvalues of the Hessian of $ \Tr[U^\dagger u\,U v]$; this determinant factor is well known and it is proportional to $\Delta(u) \Delta(v)$.
Now, due to the permutation invariance of the measure of integration for $u,v$, this sum over permutations can be performed by changing variables $v_i\rightarrow v_{\pi(i)}$ in each of the terms in the sum, so that in the end we are left with an integral over eigenvalues all lying inside a unit circle:
\begin{equation}
   \mathcal{F}(t)\simeq \oint \prod_i\frac{du_i dv_i}{u_i v_i}\,e^{N u_i v_i}\det\left(\bar{u}_j^{N+R_k-k} \right) \det\left(\bar{v}_j^{N+R_k-k} \right) \sum_{l=1}^N \frac{1}{1-t(u_l+v_l)}
\end{equation}
Now the main obstacle is that we have a pair of determinants in the integrand. To solve this issue we can expand the determinants as sums over permutations, and exploit the symmetry of the measure under index relabelling to reduce the number of sums. It is also more convenient to work with the variables $x_i= u_i v_i$ and $y_i= u_i/v_i$;
\begin{equation}
\begin{aligned}
   \mathcal{F}(t) &\simeq \frac{1}{N!}\sum_{\pi\in S_N}  (-1)^\pi \oint \prod_{i}\frac{dx_i}{x_i }\, e^{N x_i}\,\sqrt{\Bar{x}_i}^{N+R_i-i}\sqrt{\Bar{x}_{\pi(i)}}^{N+R_i-i}\\
   &\times \oint \prod_{j}\frac{dy_j}{2y_j }\sqrt{y_j}^{R_j-j}\sqrt{\Bar{y}_{j}}^{R_{\pi(j)}-\pi(j)}\, \sum_{i=1}^N\,\sum_{L=0}^\infty t^L x_i^{L/2}\left( 2+ y_i+\bar{y}_i\right)^{L/2}.
\end{aligned}
\end{equation}
 To perform the integral over $y_j$ we set $y_j=e^{2i \beta_j }$, since the coordinate $y_i$ winds around the unit circle twice. In order to get a non-zero value, all of the integrals over $y_j$ should be non-zero and since the moments only multiply by one particular value of $y_j$ for each term in the sum we can conclude that the integral is only non-zero for $\pi=id$.  After evaluating the integral over $y$ we get
 \begin{equation}
     \mathcal{F}(t) \simeq \frac{1}{N!}\oint \prod_{i}\frac{dx_i}{x_i }\, e^{N x_i}\,\Bar{x}_i^{N+R_i-i}\; \sum_i \frac{1}{\sqrt{1-4t^2 x_i}}.
 \end{equation}
This final integral is a simple Fourier integral that can be evaluated by expanding in powers of $x_i$. After normalizing $\mathcal{F}(t)$ appropriately we obtain a formula for the generating function of structure constant $C_{R R L}$:
\begin{equation}
    \mathcal{F}(t)= \sum_{L=0}^\infty t^{L}\,\frac{1^L+(-1)^L}{2\sqrt{L}} \times \left\{\binom{L}{L/2} N^{-L/2}\sum_{i=1}^N\frac{\Gamma(N+R_i-i+1)}{\Gamma(N+R_i-i-L/2+1)}\right\}.
\end{equation}
 For large representations $R_i\sim N$ with large blocks, the ratio of gamma functions can be replaced by the asymptotic expansion $\frac{\Gamma(x)}{\Gamma(x-\beta)}\simeq x^{\beta}$ as $x\rightarrow \infty$, and the sum may be replaced by an integral with $x=\left(N-i\right)/N$ and $\alpha_i=R_{k+1-i}/N$:
 \begin{equation}\label{generating function for CRRL}
   \mathcal{F}(t)= \sum_{L=0}^\infty t^L\,\frac{1^L+(-1)^L}{2\sqrt{L}} \times \left\{\binom{L}{L/2}\sum_{i=1}^k \int_{\mu_i+ \alpha_i}^{\mu_{i+1}+ \alpha_i} dx\; x^{L/2}\right\},
 \end{equation}
 here $\mu_i$ are filling fractions that measure the number of rows of size greater than or equal to $R_{k+1-i}$ in units of $N$ and we assume that there are $k$ non-zero blocks in the Young diagram for the representation $R$. Then the structure constants are:
 \begin{equation}\label{diagonalstructure}
 \begin{aligned}
   C_{R\,R\,L/2}&=\frac{1^L+(-1)^L}{2\sqrt{L}} \times \left\{\binom{L}{L/2}\int_{0}^{\infty} dx\; \rho_R(x)\times x^{L/2}\right\}\\
   &= \frac{1}{\sqrt{L}} \int dr \, r\, d\phi \, \rho_R(r) \, (z+\bar{z})^{L}, \;\;\; z= re^{i\phi/2},\\
 \end{aligned}
 \end{equation}
 where $x=r^2$ the density $\rho_R$ is defined as follows. 
\begin{figure}
\centering
\begin{subfigure}{.5\textwidth}
  \centering
  \includegraphics[width=.9\linewidth]{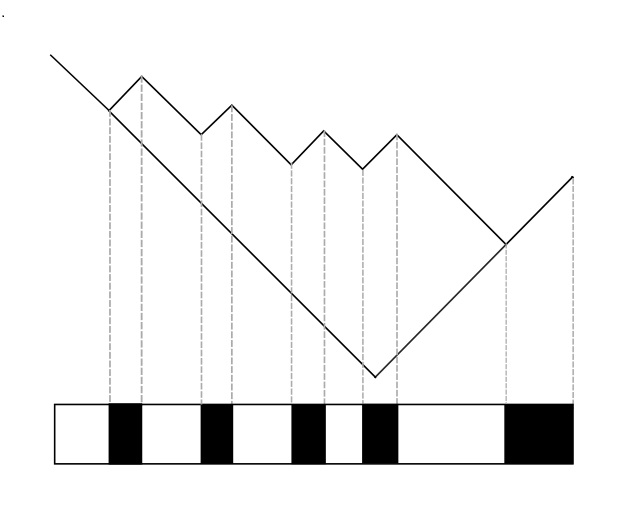}
  \caption{ }
  \label{mayadiagram}
  \label{fig:sub1}
\end{subfigure}%
\begin{subfigure}{.5\textwidth}
  \centering
  \includegraphics[width=.9\linewidth]{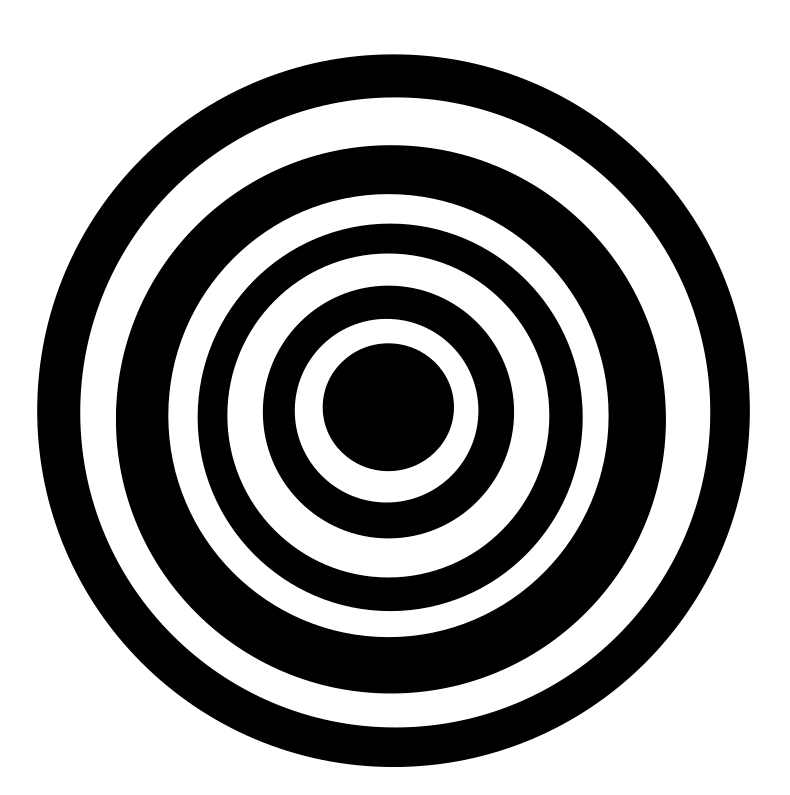}
  \caption{}
  \label{fig:sub2}
\end{subfigure}
\caption{(a) A Maya diagram associated to a large Young diagram. The rightmost positive slope edge is mapped to the Fermi sea, while negative slope edges represent large gaps of unfilled states. (b) A sketch of the LLM geometry associated to the Maya diagram.}
\label{fig:test}
\end{figure}

 Given the Young diagram we rotate it by $-3\pi/4$ radians. Then the diagram has $(k+1, k)$ edges of slopes $\mp 1$ of lengths $\mu_i$ and $\kappa_i$ for negative and positive slopes respectively. For example the length of the $l^{th}$ block is $R_i= N\sum_{i}^{k+1-i}\kappa_i=N \alpha_{i-k-1}$. We color the edges with a negative slope black, and edges with positive slopes white and unfold edges of the diagram into an infinitely long colored strip as in \ref{mayadiagram}. This strip is to be identified with a radial slice of the LLM plane \cite{deMelloKoch:2008hen}. The variable $r$ is taken to start at the right-most colored edge; this is $r=0$ in the LLM plane, and the asymptotic region is taken to be towards the left of the strip. For every black region we have $\rho_R(r)=1$, and $\rho_R(r)=0$ for every white region. Substituting this into \eqref{diagonalstructure} will reproduce the integral expression in \eqref{generating function for CRRL}. 

 This quantity can be matched precisely to the formula found in \cite{Skenderis:2007yb} for the one point point function of a chiral operator computed using holographic renormalization:
 \begin{equation}
     \langle \mathcal{O}_{S^{k \pm k}}\rangle_{LLM}= \frac{1}{\sqrt{k}}\int dr\, r d\phi \;\rho(r)\, r^{k} e^{\pm ik \phi}.
 \end{equation}
 For our background this quantity vanishes since there is a conserved $U(1)_R$ charge in the background. To match this to the expression above we take the uncharged combination $\mathcal{O}_{S^{k, +k} }+\mathcal{O}_{S^{k - k}}$ and integrate over $\phi$. This would correspond to the contribution of a spherical harmonic $Y^{L/2}\sim (z+\bar{z}+ y-\bar{y})^{L}$.
 \subsection{Off-Diagonal Structure Constant}
 To compute off-diagonal structure constants we need to change one of the representations $R$ to another representation $R'$ whose Young diagram is close to $R$. In this case most of the integrals will vanish, unless $R$ and $R'$ only differ at a single row $R_l-R'_l=k_l$. Since $R$ has large blocks, this can only happen when $R_{l-1}>R_{l}$, meaning after an edge. The only non-zero integral over the $y$ variables comes from the $y_l$ associated to the row $R'_l$, so we only get one term:
 \begin{equation}
 \begin{aligned}
     C_{R\, R+k_l\, L}&= e^{-\Delta S_{R\,R'}}N^{-L/2}\frac{1^{L-k}+(-1)^{L-k}}{2\sqrt{L}}\times\frac{L!}{(\frac{L-k}{2})!(\frac{L+k}{2})!}\,\frac{\Gamma(N+R_l+k_l/2-l +1)}{\Gamma(N+R_l+k_l/2-L/2-l +1)}\\
     &\simeq e^{-\Delta  S_{R\,R'}}\;\frac{1^{L-k}+(-1)^{L-k}}{2\sqrt{L}}\times\frac{L!}{(\frac{L-k}{2})!(\frac{L+k}{2})!}\ \alpha_l^{L/2},
 \end{aligned}
 \end{equation}
 where $ \Delta S_{R\,R'}$ is the ratio of norms $\frac{\bra{R}\ket{R}}{\sqrt{\bra{R'}\ket{R'}\bra{R}\ket{R}}}$.
 This should be interpreted as the linear response to a fluctuation localized at the edge of a particular Fermi surface within the LLM geometry. 
 \subsection{Comparing to the eigenvalue picture: fixing the number of rows}
 The method outline in this section allows us to compute expectation values of light operators in a particularly radially symmetric bubbling geometry. A natural question to address is how this connects to the eigenvalue coherent state picture. In other words, given a particular configuration of droplets, how can we determine which radially symmetry modes make up the state. Clearly a single  droplet made out of $p$ giant gravitons can only be made out of Young diagrams with $p$ rows. This is because the overlap $\bra{\tilde{\lambda}}\ket{\lambda}$ has a character expansion, and setting $N-p$ of the eigenvalues to zero makes characters associated to representations with more than $p$ rows vanish. To project to a particular diagram made out of $p$ rows we repeat the same trick where we integrate the remaining eigenvalues over a unit circle with an appropriate measure. After regrouping the integration variables we will end up with a pair integrals over $U(p)$:
 \begin{equation}
 \begin{aligned}
      \bra{R',p}\ket{R,p}&\propto \int d\mu(s^\dagger) d\mu(s) \, \det\left(\mathbb{I}- s^\dagger s\right)^{N-2p}
     \int_{U(p)} d U \int_{U(p)} dV \, S_{R}(U^\dagger) S_{R'}(V^\dagger) e^{N\Tr[U s V s^\dagger]}\\
     &\propto \frac{\delta_{R R'}}{Dim_N(R)^2}\int d\mu(s^\dagger) d\mu(s) \, \det\left(\mathbb{I}- s^\dagger s\right)^{N-2p} S_R(s^\dagger s)\\
     &\propto \frac{\delta_{R R'}}{Dim_N(R)^2}\int_{[0,1]^p} \prod_{i=1}^p dx_i\,\Delta_p(x) \,\left(1- x_i \right)^{N-2p} x_i^{p+R_i-i}.
 \end{aligned}
 \end{equation}
For generic (non-rectangular) diagrams, the integral does not have a simple solution. For large diagrams we can use the saddle point approximation to find the density of eigenvalues $x_i$. The procedure to evaluate this class of integrals was outlined in \cite{Halmagyi:2007rw} and also \cite{Okuda:2008px, Gomis:2006mv, Gomis:2008qa}. For a Young diagram with $k$ large blocks with $n_a$ rows, we split the variables $x_i$ into $k$ groups of size $n_a$, $x_i=(x_{1}^{(1)}, \dots, x_{n_1}^{(1)},x_{(1)}^{2}, \dots,  x_{n_2}^{2}, \dots, x_{1}^{k}, \dots, x_{n_k}^{(k)} )$. Then we use the fact that the integration variables are invariant under permutations to rewrite 
\begin{equation}
\begin{aligned}
   &\int \prod_{a=1}^k \prod_{i_a=1}^{n_a} dx_{i_a}^{(a)} \left( x_{i_a}^{(a)}\right)^{R_a-N_a}      \left(x_{i_a}^{(a)}\right)^{n_a-i_a}\\
   &\int \prod_{a=1}^k \prod_{i_a=1}^{n_a}\, dx_{i_a}^{(a)}\, \times \frac{1}{n_a!}\sum_{\pi\in S_{n_a}} \left( x_{i_a}^{(a)}\right)^{R_a-N_a}     \left(x_{\pi(i_a)}^{(a)}\right)^{n_a-i_a}\\
   &\int \prod_{a=1}^k \prod_{i_a=1}^{n_a} dx_{i_a}^{(a)} \left( x_{i_a}^{(a)}\right)^{R_a-N_a}      \Delta_{n_a}(x^{(a)}),
\end{aligned}
\end{equation}
where $N_a$ is the partial sum $\sum_{b\leq a} n_b$ and $N_1=0$. Putting it all together, we are left with a multi-cut matrix model:
\begin{equation}
    \mathcal{Z}(R)= \int_{[0,1]^p} \prod_{a=1}^k \,\prod_{i_a}^{n_a} dx_{i_a}^{(a)}\, \Delta_{n_a}\left(x^{(a)}\right)^2\,\left(1-x_{i_a}^{(a)} \right)^{N-2p}  \left( x_{i_a}^{(a)}\right)^{R_a-N_a}\times \prod_{c>b}\,\prod_{i_b, j_c} \left(x^{(c)}_{j_c}- x^{(b)}_{i_b}\right)
\end{equation}
A matrix model quite similar to this one studied in \cite{Halmagyi:2007rw} for computing vevs of giant Wilson loops in Chern-Simons theory on Lens spaces $S^3/\mathbb{Z}_p$. To make the analogy more precise we can change variables to $x_i=e^{-u_i}$ after which the partition function becomes:
\begin{equation}
\begin{aligned}
    \mathcal{Z}(R)&= \int  \prod_{a=1}^{k}\prod_{i} du_{i}^{(a)} \prod_{j<i}  \left(2 \sinh \frac{u_j^{(a)}-u_i^{(a)}}{2}\right)^2 \left(2 \sinh \frac{u_i^{(a)}}{2}\,\right)^{N-2p}\\
    &\times e^{-(L_a+N/2-p) \,u_i^{(a)}}\prod_{b>a}\prod_{i,j} \left(2\sinh \frac{u_j^{(b)}- u_i^{(a)}}{2}\right).\\
    L_a&= R_a-\frac{1}{2}\sum_{b=1}^{a-1} n_b+ \frac{1}{2}\sum_{b>a} n_b
\end{aligned}
\end{equation}
The only difference between this and the matrix model studied in \cite{Halmagyi:2007rw} is that the Gaussian term is replaced by $(1-e^{-u_i^{(a)}})^{N-2p}$, and $u$ differs by a sign. In this representation the saddle point analysis is quite straightforward. The equations of motion for the eigenvalues $u_i^{(a)}$ take the form:
\begin{equation}\label{saddle point YT}
    \frac{1}{N} \; \sum_{j\neq i}\coth \frac{u_{j}^{(a)}-u_i^{(a)}}{2} +\frac{1}{2N}\sum_{b\neq a, \,l} \coth\frac{u^{(b)}_l-u_i^{(a)}}{2}= \frac{(1-2p/N)}{2}\left(\coth\frac{u^{a}_i}{2}-1\right)- L_a/N.
\end{equation}
The resolvents for this type of problem can be taken to be of the form $\omega^{(a)}(z)=\frac{1}{N}\sum_{i}\coth \frac{z-u_{i}^{(a)}}{2}$, and the total resolvent is $\omega(z)= \sum_{b=1}^k \omega^{(b)}(z) $. At large $N$ the eigenvalues condense into $k$ branch cuts and the saddle point equation on the $a^{th}$ cut reads
\begin{equation}
\omega^{(a)}(z+i\,0)+\omega^{(a)}(z-i\,0)+\sum_{b\neq a}\omega^{(b)}(z)= \frac{(1-2p/N)}{2}\left(\coth\frac{u^{(a)}_i}{2}-1\right)- L_a/N,
\end{equation}
or equivalently
\begin{equation}\label{RH}
\omega(z\pm i\,0)=-\omega^{(a)}_{\mp}(z)+ \frac{(1-2p/N)}{2}\left(\coth\frac{z}{2}-1\right)- L_a/N.
\end{equation}
This equation defines a Riemann-Hilbert problem for the total resolvent $\omega(z)$. Notice that for large enough values of $z$, the potential rapidly approches a constant value of $-L_a/N$, and for small values of $z$ there is exponential barrier pushing the eigenvalues away from $z=0$. This means that the eigenvalues will sit far from $z=0$, and will be uniformly distributed along the cut. To solve \eqref{RH} we need to find a function of the resolvents that is regular for $\Re(z)>0$. Following \cite{Halmagyi:2007rw} we can define a set of complex variables 
\begin{equation}
\begin{aligned}
    X_0&= e^{-(1-2p/N)\left(\coth\frac{z}{2}-1\right)/2}  e^{\omega}=W\,e^{\omega}\\
    X_a&= e^{-L_a/N} e^{-\omega_a};
\end{aligned}
\end{equation}
then the equations \eqref{RH} are equivalent to a monodromy condition for the $X_I$ as we around each of the cuts. The solution to this problem is given by a polynomial of degree $(k+1)$, $f(Y,W)$, with roots at the $X_I$ and the spectral curve is the zero locus of this polynomial. The precise eigenvalue distribution is then found by solving $f(Y,W)=0$ for $Y$ and taking the discontinuity of the resolvent $\omega\propto \log Y$ at each cut.

In the thermodynamic limit, the branch cuts become widely separated and the saddle point equations simply significantly. Substituting in the ansatz $u^{(b)}_j- u^{(a)}_i\gg 1$ in \eqref{saddle point YT} turn the terms of the form $\coth \frac{u^{(b)}_j- u^{(a)}_i}{2}$ into a constant. The interaction term between eigenvalues in the same cut becomes a step function for large $u^{(a)}$ beginning at the first eigenvalue on the cut which given an equation for each of the eigenvalues on a given cut:
\begin{equation}
\begin{aligned}\label{distribution eq}
    \frac{(1-2p/N)}{2}\left(\coth\frac{u^{a}_k}{2}-1\right)= \frac{1}{N}\left( L_a- (n^{(a)}-k) + \frac{1}{2}\sum_{b>a}n_b\right)\Rightarrow\\
 2g_s \left(\frac{N-2p}{1-x_k}\right)= g_s\left[R_a -\sum_{b=1}^a n_b+ \sum_{c>a}n_c+ \left(2k +\sum_{b>a} n_b+ R_a\right)+2\right].
\end{aligned}
\end{equation}
Clearly the eigenvalues $\lambda_i= 1/(1-x_i)$ are uniformly distributed along each cut, and they are quantized in units of $2g_s= 1/(N-2p)$. So again we see that this is indeed the correct parameter controlling the fluctuations around the saddle. These equations are also exactly analogous to the saddle point equations for the single giant graviton case,
\begin{equation}
    \frac{1}{1-x_k}= |\lambda_k|^2=r_k^2 
\end{equation}
and the distribution $\rho_R(r)$ for $r_k$ is the same distribution we found before. 
\subsubsection{Computing correlators: reduced matrix elements}
To compute correlation functions of single trace operators between a set of states $\ket{R,p}, \ket{R',p}$ we need to perform a series of elaborate coordinate substitutions to simplify the matrix integral calculations. First, the states are obtained by a certain projection of the reduced coherent generating function 
\begin{equation}
\begin{aligned}
     \ket{R,p}&\propto\int_{U(p)} dU_1\int dV_1 dV_1^\dagger \; e^{\tr_p[U_1 V_1^\dagger Z V_1]} S_R(U_1^\dagger) \ket{0}\\
     \bra{R',p}&\propto\int_{U(p)} dU_2\int dV_2 dV_2^\dagger \;\bra{0} e^{\tr_p[U_2 V_2^\dagger \bar{Z} V_2^\dagger]} S_{R'}(U_2^\dagger).
\end{aligned}
\end{equation}
If we insert a half BPS single trace operator between these states, the scalar fields $Z$ and $\Bar{Z}$ inside the trace are traded by complex valued matrices:
\begin{equation}
\begin{aligned}
    Z\rightarrow V_2U_2 V_2^\dagger \\
     \bar{Z}\rightarrow V_1 U_1 V_1^\dagger .
\end{aligned}
\end{equation}
We can then perform a color-flavor transformation to express any trace of a power of $Z+\bar{Z}$ (over $N$ color indices) into a trace over $2p\times 2p$ matrices.  Then after performing all of the contractions between the two exponentials we end up with a pair of unitary integrals over $U_{1,2}$ and an integral over the $p\times p$ `radial' matrix $\sigma =V_1^\dagger V_2$.

\begin{equation}
\begin{aligned}
&\bra{R',p} \Tr\left[ \left(Z+\bar{Z}+ Y-\Bar{Y} \right)^L\right] \ket{R,p}=\\
&\int dU_1 dU_2 \,d\sigma^\dagger d\sigma \Bigg\{ \det\left(1-\sigma^\dagger\sigma \right)^{N-2p}  e^{\tr_p\left[U_1 \sigma U_2 \sigma^\dagger\right]}\, S_{R}(U_1^\dagger) S_{R}(U_2^\dagger)\\& \times \tr_{2p}\left[ \begin{pmatrix}
    U_1& \sqrt{U_1}\, \sigma \sqrt{U_2}\\ \sqrt{U_2}\,\sigma^\dagger \sqrt{U_1}& U_2
\end{pmatrix}^L\right] \Bigg\},    
\end{aligned}
\end{equation}
and here $\sqrt{U_{1,2}}$ refers to the unitary matrix whose eigenvalues are square roots of the eigenvalues of $U_{1,2}$. To solve this integral at large $N$ we will diagonalize $U_{1,2}$ and perform a singular value decomposition of $\sigma$.
\begin{equation}
    \begin{aligned}
       \sigma&= U_{L}\, s \,U_{R}\\
       U_{1}&= \mathcal{U}^\dagger e^{i \boldsymbol{\alpha}}\, \mathcal{U}\\
         U_{2}&= \mathcal{V}^\dagger e^{i \boldsymbol{\beta}}\, \mathcal{V}.
    \end{aligned}
\end{equation}
Although this at first glance looks daunting one should realize that the translational invariance of the Haar measure allows us to reabsorb most of the unitary integrals into the integration over $U_L$ and $U_R$. This reduces the calculation to a pair of integrals involving simple complex exponential associated to the eigenvalues of $U_{1,2}$ which encode the angular dependence of the correlator, one integral over the radial eigenvalues, and a pair of difficult unitary integrals over $U_{L,R}$. First we address the integration over $U_{L}$ and $U_{R}$. At large $N$ we can perform a saddle point approximation for this integral; since the only term that is relevant for this is the exponential we are left with the task of finding the critical points of the following function:
\begin{equation} \label{four matrix potential}
    S(U_L, U_R)= \tr_{p}\left[U_L^\dagger e^{i \boldsymbol{\alpha}}\, U_L \, s\, U_R\, e^{i \boldsymbol{\beta}} U_{R}^\dagger \,s^\dagger \right].
\end{equation}
The critical points of this function are given by the solutions to a pair of matrix equations
\begin{equation}
    [s^\dagger U_L^\dagger e^{i \boldsymbol{\alpha}}\, U_L \, s\, ,U_R\, e^{i \boldsymbol{\beta}} U_{R}^\dagger]=[ s\, U_R\, e^{i \boldsymbol{\beta}} U_{R}^\dagger \,s^\dagger,U_L^\dagger e^{i \boldsymbol{\alpha}}\, U_L ]=0.
\end{equation}
These equations essentially imply that these two pairs of matrices are simultaneously diagonalizable. For example, the first equation is unitarily equivalent to the condition that $e^{i \boldsymbol{\beta}}$ and $U_R^\dagger s^\dagger U_L^\dagger e^{i \boldsymbol{\alpha}}\, U_L \, s\,U_R$ are both diagonal in the same basis. The second equation gives a similar condition for $ e^{i \boldsymbol{\alpha}}$ and $U_L\, s\, U_R\, e^{i \boldsymbol{\beta}} U_{R}^\dagger \,s^\dagger\, U_L^\dagger$. But since $ e^{i \boldsymbol{\alpha}}$, $e^{i \boldsymbol{\beta}}$ and $s$ are all diagonal the only way that this can be achieved is if $U_{L,R}$ are permutation matrices. This is clear because making the ansatz $U_R= U_{\pi}$  in the first equation for some permutation matrix $U_\pi$ immediately forces $U_L$ to be a permutation matrix and vice versa. This means that we have one critical point for every pair of permutations $\pi, \tau$ in $S_p$, which act on  $ e^{i \boldsymbol{\alpha}}$ and  $ e^{i \boldsymbol{\beta}}$ by permuting their eigenvalues independently from each other. To get the correct answer we should also compute the one-loop determinant around each of the saddles. This boils down to computing the Hessian of \eqref{four matrix potential}, which is the same for each each critical point up to a sign associated to the determinant of the permutation. To quadratic order $S(e^{M}, e^{N})$ is
\begin{equation}
\begin{aligned}
  S(m,n)&\simeq\tr_p[ e^{i \boldsymbol{\alpha}} \,s \, e^{i \boldsymbol{\beta}}\, s^\dagger] -\frac{1}{2}\sum_{i,j}\Bigg\{ (e^{i\alpha_i}-e^{i\alpha_j})(s_i s_i^*\,e^{i\beta_i}-s_j s_j^* e^{i\beta_j})|m_{ij}|^2 \\
  &+ (e^{i\beta_i}-e^{i\beta_j})(s_i s_i^*\,e^{i\alpha_i}-s_j s_j^* e^{i\alpha_j})|n_{ij}|^2 \\
  &-(e^{i\alpha_i}-e^{i\alpha_j})(e^{i\beta_i}-e^{i\beta_j})s_is_j^* \left(m_{ij}n_{ij}^*+n_{ij}m_{ij}^*\right)\Bigg\}+\dots,
\end{aligned}
\end{equation}
so the one-loop factor becomes simply
\begin{equation}
    \det\left( \partial_{ij}\partial_{kl} S(m,n)\right)= \Delta_p\left( e^{i \boldsymbol{\alpha}}\right) \Delta_p\left( e^{i \boldsymbol{\beta}}\right) \Delta_p\left(s^\dagger s\right)\Delta_p\left( e^{i (\boldsymbol{\alpha}+\boldsymbol{\beta})} s^\dagger s\right).
\end{equation}
Each saddle point is weighted by $\det(\tau \pi)$, so after a coordinate transformation every term in the sum will give the same value. As before, the one-loop factor when combined with the denominators of the determinantal expressions for the Schur polynomials  will cancel the Vandermonde determinants in the integration measure of $\alpha_i$ and $\beta_i$, which makes the integrals over the angular variables straightforward

\begin{equation}
\begin{aligned}\label{reduced correlator}
&\bra{R',p} \Tr\left[ \left(Z+\bar{Z}+ Y-\Bar{Y} \right)^L\right] \ket{R,p}=\\
&\int \frac{d\alpha_i}{2\pi } \frac{d\beta_i}{2\pi }\int_{[-1,1]^p} ds_i\,\Bigg\{ \frac{\Delta_p(s^\dagger s )}{\Delta_p( e^{i (\boldsymbol{\alpha}+\boldsymbol{\beta})}s^\dagger s)} \left(1-|s_i|^2 \right)^{N-2p}  \exp \left(e^{i( \alpha_i+\beta_i)}|s_i|^2\right)\\ &\det_{l,k}\left(e^{-i(p+R_l-l)\alpha_k}\right) \det_{l,k}\left(e^{-i(p+R'_l-l)\beta_k}  \right)  \times\binom{L}{L/2} \tr_{p}\left[  (s^\dagger s)^{L/2}\, e^{i(\frac{\boldsymbol{\alpha}+\boldsymbol{\beta}}{2})L} \right] \Bigg\}.  
\end{aligned}
\end{equation}
For light operators and diagonal structure constants we can simply interchange $(Z+\bar{Z})^{L}$ with $\binom{L}{L/2}(Z\Bar{Z})^{L/2}$, which is simpler to work with. The integration over $\alpha_i-\beta_i$ will again force the pair of sums over permutations coming from the two determinants to collapse to a single sum, and the integral over $\alpha_i+\beta_i$ cancels the factor of $(s_k s_k^*)^{L/2}$. Finally the integral over $s_i$ factors out completely and the correlator comes only from the integrals over the angular variables:
\begin{equation}
\begin{aligned}
&\bra{R,p} \Tr\left[ \left(Z+\bar{Z}+ Y-\Bar{Y} \right)^L\right] \ket{R,p}\\&=\mathcal{Z}(R) \sum_i \frac{1^L+(-1)^L}{2} \times \left\{\binom{L}{L/2}N^{-L/2}\sum_{i=1}^p\frac{\Gamma(p+R_i-i+1)}{\Gamma(p+R_i-i-L/2+1)}\right\}.
\end{aligned}
\end{equation}
This is the same answer as \eqref{diagonalstructure} except that we are missing one term coming from the droplet made out of $N-p$ spectator branes. The reason that we miss this term in this calculation is that the reduced generating functions project out all Young diagrams with more than $p$ rows, which essentially freezes $N-p$ of the branes that make up the background. This agrees with our interpretation of the off-diagonal structure constant as the linear response of a particular Fermi sea level. Since we projected out all the contributions from Young diagrams with more than $p$ rows, ripples of the first Fermi sea surface are projected out from this computation.  
\section{Discussion and Future Directions}
\renewcommand{\theequation}{5.\arabic{equation}}
\setcounter{equation}{0}
In this paper we introduced techniques for dealing with large BPS operators in $\mathcal{N}=4$ SYM theory and revisited the computation of one-point functions in the background of states corresponding to the bubbling geometries of \cite{Lin:2004nb}. Our method is based on the semi-classical techniques introduced in \cite{Berenstein:2022srd, Jiang:2019xdz, Chen:2019gsb} and further developed in \cite{Holguin:2022zii, Gaiotto:2021xce, Holguin:2022drf, Lin:2022gbu, Lin:2022wdr} and provides an independent derivation of the results of \cite{Takayama:2005yq}. One advantage of our methods is that they do not rely on diagonalizing any the field operators of the theory, or performing any kind of consistent truncation of the model, which makes weak coupling computation of non-BPS observables possible. We also fleshed out the relation between coherent states and characters by providing an explicit integral transform between both pictures. This gives somewhat complementary methods that can be used to compute correlators of operators describing somewhat generic LLM backgrounds. Although our results are expected and perhaps unsurprising, we hope that the techniques developed here serve as a starting point for performing a systematic large $N$ expansion for large operators. 

One question worth asking is whether we learned anything about the statistics of (BPS) OPE coefficients in $\mathcal{N}=4$ SYM. We are hesitant to claim that our results display any kind of chaotic behavior predicted by the eigenstate thermalization hypothesis \cite{Srednicki:1994mfb}, since half-BPS operators in $\mathcal{N}=4$ SYM are completely captured by a free theory. We instead attribute the appearance of random matrix statistics to the averaging necessary to describe large semi-classical states, and to effects due to the large charge limit. In fact our computations suggest that true structure constants (as opposed to fixed charged three point functions) only receive contributions from a single term, and that the would be distribution of eigenvalues in these cases are essentially constant distributions. On the other hand, correlators involving large operators that break the $R$-symmetry spontaneously but with fixed charge are generically given by complicated averages. In the large $N$ limit with the charges of the operators scaling with $N^2$ this leads to the appearance of random matrix behavior in the OPE coefficients. Interestingly similar distributions were found to emerge from the large charge limit of extremal correlators in rank one SCFTs \cite{Grassi:2019txd}. 

We conclude by commenting on some future directions of work. 
\subsection*{Holographic computation of off-diagonal three-point functions}
Our methods allow us to make predictions for the value of the off-diagonal structure constant between two LLM geometries. This quantity is only non-zero when the two geometries differ by a small fluctuations. It would be ideal to try develop semi-classical techniques for computing such things using the gravitational path integral. Since states are highly degenerate (there is one for each Young diagram!), we expect that there is a set of commuting charges that differentiates between different states in gravity. These charges would correspond to some kind of higher spin asymptotic symmetries of LLM solutions. Then, computing off-diagonal three-point functions would correspond to dressing the semi-classical saddle by a wavefunction charged with the appropriate asymptotic charge as suggested by \cite{Yang:2021kot}. This would be a very simple toy model of the soft hair proposal \cite{Hawking:2016msc} in the sense that one is able to probe very precise details of the interior of the geometry from simple boundary manipulations. Such a technique would also have to go far beyond our current methods of holographic renormalization \cite{Bianchi:2001kw}. We take the fact that the formulas for one-point functions in LLM backgrounds are quite simple as indicative of the existence of a different method for computing holographic one point functions in them. Perhaps a careful WKB analysis \cite{Grinberg:2020fdj} might be able to reproduce the result for operators of arbitrary charge without the need for a non-linear Kaluza-Klein reduction.
\subsection*{Three Heavy Operators}
One obvious extension of our calculations would be to consider the structure constants of three really heavy operators half-BPS operators. There are in principle no conceptual obstructions for performing such computations, since we only have to include an additional exponential generating function. Technically speaking the saddle point analysis is more involved and we expect the form of the structure constants to be much richer.  More precisely many of the simplifications that occur for the one-point functions of single trace operators are basically due to the fact that for small operators the term in the tensor decomposition only show up with multiplicity one. This is why we are able to reduce the number of sums in \eqref{reduced correlator}. For example one can insert a particular Schur polynomial of $Z+\bar{Z} + Y-\bar{Y}$ between two coherent states. In this case we can commute the coherent states past the Schur polynomial resulting in a matrix model generalizing \eqref{normp}.  Obtaining an approximate formula even for extremal three point functions of very large operators would be a rather non-trivial since would encode information about the statistics of Littlewood-Richardson coefficients. On the other hand, such correlators predict the existence of supergravity solutions which interpolate between different LLM geometries \cite{Jiang:2019xdz}. Since these types of three point functions are protected, we expect that one should be able to match both results precisely in the large $N$ limit, so obtaining an approximate form for such correlators might give some intuition about how to construct such geometries. Perhaps a simpler problem is to consider correlators of two LLM geometries and a giant graviton, or between three giant gravitons. These quantities give predictions for giant graviton nucleation amplitudes. 
\subsection*{Extremal Correlators}

Another immediate generalization would be to study higher point extremal correlators involving various combinations of single trace, giant graviton, and LLM geometries. These correlators are also protected, so we can hope to be able to perform a holographic check for these quantities as well. In practice this will likely involve a very careful group averaging procedure to be able to overcome any possible ambiguities that often arise in extremal correlators. For instance one can conceivably study correlators in which many branes nucleate into a bubbling geometry, or where large droplets all fuse into a single droplet, or a large droplet splits into smaller ones. Such processes are somewhat reminiscent of baby universe creation and annihilation processes. Whether such a geometric picture can be realized from the bulk point of view is unclear.
\subsection*{Worldsheet and spin chain interpretation of one-point functions in LLM background}
One particularly puzzling issue is to interpret the result of the computation of a three point function of a non-protected single trace primary and two very large half-BPS operators. For large operators of dimension of order $N$ the interpretation as a worldsheet $g$-function was advocated in \cite{Jiang:2019xdz}, and various checks were performed. This makes sense since operators of dimension of order $N$ can change the boundary conditions of the string worldsheet in the bulk. For operators of dimension of order $N^2$ this interpretation is inadequate, since we expect that the correct description is instead in terms of a string moving in an LLM background, rather than simply an open string ending on a stack of branes. From the worldsheet point of view one would expect that the worldsheet CFT flows as the background is deformed. Understanding what this would mean from the point of view of the spin chain picture would be quite interesting. 
\subsection*{$\frac{1}{4}$, $\frac{1}{8}$, and $\frac{1}{16}$ BPS operators}
A more long term goal would be to understand the systematics of less supersymmetric BPS operators in the $\mathcal{N}=4$ SYM theory. In principle introducing additional scalar matrices in the exponentials gives a way of generating $\frac{1}{4}$ and $\frac{1}{8}$ BPS \cite{Berenstein:2022srd, Chen:2019gsb, Carlson:2022dot}. This was carried out for rank two $\frac{1}{4}$ BPS operators in \cite{Carlson:2022dot}. Additionally, the generating functions introduced in \cite{Berenstein:2022srd} contain all bosonic $\frac{1}{8}$ operators. The main difficulty lies in constructing explicit expressions for restricted Schur polynomials with which one can project to particular BPS operators of the weakly coupled theory. In other words, finding the integration rules for the analog of the Schur polynomials in \eqref{projector} with multiple matrices would be a big step in this program. This problem is even more stark for $\frac{1}{16}$ BPS operators where one seems needs to need to introduce an infinite number of matrices corresponding to covariant derivatives of the scalar fields. In analogy to the construction in \cite{Berenstein:2022srd}, one can generate $\frac{1}{16}$ BPS states by exponentiating the so-called $\frac{1}{16}$ BPS letter introduced in \cite{Chang:2013fba}. Understanding precisely what kind of excitations are relevant for studying supersymmetric black holes would be a first step towards a boundary derivation of the results of \cite{Boruch:2022tno}.
\subsection*{Twisted Holography}
A natural setting where our techniques can be readily applied is in the context of the  Twisted Holography program. For instance sphere giant gravitons and non-conformal vacua have already been studied \cite{ Budzik:2021fyh, Budzik:2022hcd} and their geometric picture is quite clear. Extending this to include the analog of AdS giant gravitons and more generic Schur polynomial operators seems like a straightforward task. It would be nice to develop the more conventional view of giant gravitons wrapping compact cycles on the deformed conifold $SL(2, \mathbb{C})$ in the B-model by developing a global version of Twisted Holography, in analogy to global AdS holography. This should be related to studying the chiral algebra on $\mathbb{P}^1$ instead of $\mathbb{C}$. The point of this exercise is that when we insert an operator associated to a Young diagram $R$, there is a well-studied recipe for producing a spectral curve, and hence a bubbling Calabi-Yau manifold \cite{Halmagyi:2007rw}. This story is well-understood for the $A$-model on the deformed conifold, so it is not implausible that a similar story exists for the $B$-model. This might help bridge the conceptual gap about the relation between the full physical holography and the twisted version.

\subsection*{Semiclassics and Large $N$}
Inevitably one will need to include $g^2 N$ corrections in a systematic way when dealing with non-protected operators. A good place to start would be consider simpler models where the techniques developed in \cite{Hellerman:2015nra, Badel:2019oxl, Cuomo:2021ygt} can be combined with the large $N$ expansion. Performing a near BPS expansion seems like a natural starting point since one expects the coupling constant to be enhanced by additional kinematic effects \cite{Berenstein:2002jq} making a reliable extrapolation to strong coupling a possibility for certain observables. A good target would to understand the correlation functions of more complicated `baryonic' operators in the Wilson-Fisher fixed point of the $O(N)$ model \cite{giombi2020large} in the large $N$ limit;

\begin{equation}
    \mathcal{B}\sim S_{i_1, \dots , i_N} \, \phi^{i_1}\dots \phi^{i_{N}}
\end{equation}
 where $S$ is a symmetric tensor. The simplest example of these operators where studied in \cite{giombi2020large}, although it would be interesting to extend these kinds of results to operators associated to other representations of $O(N)$. These kind of operators can also be exponentiated with the help of real gaussian integrals over Grassmannians, so the main difficulty would be to study the gap equations in the presence of these baryons.  
\renewcommand{\theequation}{1.\arabic{equation}}
\setcounter{equation}{0}

\section*{Acknowledgments}
We would like to thank David Berenstein, Sanjaye Ramgoolam, and Antal Jevicki for discussions. We are pleased to thank Kwinten Fransen, Shota Komatsu, Robert de Mello Koch, and  Wayne Weng for comments on an earlier draft. A.H. is supported in part by funds from the University
of California.

\renewcommand{\theequation}{A.\arabic{equation}}
\setcounter{equation}{0}


\bibliographystyle{unsrt}
\bibliography{references}
\end{document}